\documentclass[aps,reprint]{revtex4-1}
\usepackage[english]{babel}
\usepackage[utf8]{inputenc}
\usepackage[colorinlistoftodos, color=green!40, prependcaption]{todonotes}
\usepackage{amsthm}
\usepackage{amsmath}
\usepackage{amssymb}
\usepackage{mathtools}
\usepackage{physics}
\usepackage{xcolor}
\usepackage{graphicx}
\usepackage[left=20mm,right=20mm,top=25mm,columnsep=15pt]{geometry} 
\usepackage{comment}
\usepackage{adjustbox}
\usepackage{placeins}
\usepackage[T1]{fontenc}
\usepackage{lipsum}
\usepackage{csquotes}
\newcommand{\tg}[1]{\textcolor{black}{#1}}

\usepackage[pdftex, pdftitle={Article}, pdfauthor={Author}]{hyperref} % For hyperlinks in the PDF
\begin{document}
%\title{Rheo-Echo-XPCS and the yielding of a colloidal fractal gel}
\title{Tracking microscopic irreversibility during yielding of a colloidal fractal gel with Rheo-Echo-XPCS}

\author{William Chèvremont$^1$, Julien Bauland$^2$, Emmeline Brassac$^2$,  Gonzalo Sanchez Vera$^3$, Stefano Aime$^3$, Frédéric Pignon$^4$ and Thomas Gibaud$^{2,5}$}
    \email[Corresponding author, ]{thomas.gibaud@ens-lyon.fr}% Your name
    \affiliation{$^1$ESRF, The European Synchrotron, 71 Avenue des Martyrs, CS40220, 38043 Grenoble Cedex 9, France}
    \affiliation{$^2$ENSL, CNRS, Laboratoire de Physique, F-69342 Lyon, France}
    \affiliation{$^3$C3M, ESPCI Paris, Université PSL, CNRS, 75005 Paris, France}
    \affiliation{$^4$Univ. Grenoble Alpes, CNRS, Grenoble INP, LRP, F-38000 Grenoble, France}
    \affiliation{$^5$Department of Polymer Engineering, IPC, University of Minho, Guimarães, 4804-533 Portugal}
%%%%%%%%%%%%%%%%%%%%%%%%%%%%%%%%%%%%%

\date{\today}
%%%%%%%%%%%%%%%%%%%%%%%%%%%%%%%%%%%%
\begin{abstract}
Understanding how microscopic structural dynamics relate to macroscopic mechanical response during yielding remains a central challenge in soft matter physics. Here, we introduce rheo-echo X-ray photon correlation spectroscopy (rheo-echo-XPCS) with nonlinear acquisition synchronized to oscillatory shear, enabling direct measurement of irreversible nanoscale dynamics under strain amplitude control. Applying this to a carbon black colloidal fractal gel, we resolve time-periodic echoes in the vorticity-direction intensity autocorrelation function whose decay encodes non-affine structural rearrangements. We find: (i)~\tg{ballistic-like} decorrelation with $\tau \propto q^{-1}$ at all strains, where the decorrelation velocity $v_\tau = 1/\langle q\tau \rangle$ scales linearly with the loss tangent $\tan\delta = G''/G'$, establishing $\tan\delta$ as a \tg{direct macroscopic signature of the rate of irreversible structural decorrelation}; (ii)~functional form continuous evolution from compressed exponential ($\alpha \simeq 1.5$) at low strain—consistent with three-dimensional dipolar strain fields in the intact network—to stretched exponential ($\alpha \simeq 0.5$) at high strain, reflecting a dimensional reduction from $d_f = 3$ to $d_f = 1$ as stress transmission shifts from bulk to quasi-one-dimensional filamentary backbones during network fragmentation.
\end{abstract}
%%%%%%%%%%%%%%%%%%%%%%%%%%%%%%%%%%%%
\begin{abstract}
%%%%%%%%%%%%%%%%%%%%%%%%%%%%%%%%%%%%
Understanding how microscopic structural dynamics relate to macroscopic mechanical response during yielding remains a central challenge in soft matter physics. Here, we introduce rheo-echo X-ray photon correlation spectroscopy (rheo-echo-XPCS) with nonlinear acquisition synchronized to oscillatory shear, enabling direct measurement of irreversible nanoscale dynamics under controlled strain amplitude.
Applying this approach to a carbon black colloidal fractal gel, we resolve time-periodic echoes in the vorticity-direction intensity autocorrelation function, whose decay encodes cycle-to-cycle non-affine structural rearrangements spanning the linear viscoelastic regime, nonlinear deformation, and the yielding regime beyond $\gamma_y$.
We find: (i) ballistic-like decorrelation with $\tau \propto q^{-1}$ at all strain amplitudes, where the decorrelation rate $v_\tau = 1/\langle q \tau \rangle$ scales linearly with the loss tangent $\tan\delta = G''/G'$, establishing $\tan\delta$ as a macroscopic measure of the rate of irreversible structural decorrelation; (ii) a continuous evolution of the relaxation function from compressed exponential ($\alpha \simeq 1.5$) at low strain—consistent with three-dimensional dipolar strain fields in the intact network—to stretched exponential ($\alpha \simeq 0.5$) at high strain, reflecting a dimensional reduction from $d_f = 3$ to $d_f = 1$ as stress transmission shifts from bulk to quasi-one-dimensional filamentary backbones during network fragmentation.
\end{abstract}
%%%%%%%%%%%%%%%%%%%%%%%%%%%%%%%%%%%%

%%%%%%%%%%%

\maketitle %\maketitle must follow title, authors, abstract and \pacs

{\section{Introduction}}
%%%%%%%%%%%%%%%%%%%%%%%%%%%%%%%%%%%%
\tg{Yielding induced by shear stress or strain in soft solids, the progressive transition from predominantly solid state to predominantly fluid state}, is a central problem in soft matter physics and underpins a wide range of industrial and biological processes~\cite{bonn2017,gibaud2020}. Macroscopic rheology identifies yielding through stress overshoots, moduli crossovers, or loss of linearity, yet such bulk signatures provide only indirect access to the microscopic processes responsible for irreversibility. %Elucidating how nanoscale structural rearrangements accumulate and organize to produce macroscopic flow therefore requires direct, time-resolved probes that couple structural dynamics to mechanical response throughout the yielding process.
While significant progress has been made in understanding the yielding behavior of colloidal glasses~\cite{knowlton2014,keim2014,rogers2018, aime2023}
colloidal gels pose a qualitatively different challenge. Unlike dense glasses, gels are sparse, hierarchical networks in which attractive colloids assemble into fractal clusters that percolate to form stress-bearing backbones. This open architecture raises fundamental questions: does yielding proceed via mechanisms analogous to those in glasses, or through distinct pathways involving bond rupture, cluster rearrangements, and network fragmentation? Recent studies demonstrate that yielding in gels is progressive and history dependent. Rheological measurements have shown strong protocol dependence of yield points~\cite{dinkgreve2016}, while constant-load experiments~\cite{aime2018} revealed long induction periods during which bulk rheology remains linear even as microscopic rearrangements accelerate prior to failure. Continuum modeling~\cite{kamani2021} further demonstrated that plastic deformation develops below the nominal yield stress, and recent work~\cite{keane2025} showed that the magnitude of unrecoverable strain is quantitatively predicted by tan~$\delta$ measured in the linear regime. Dynamic scattering studies~\cite{donley2020, kamani2025} corroborated this picture by linking yielding to a continuous evolution from recoverable to irreversible dynamics.

Despite these advances, direct microscopic measurements connecting structural relaxation to network fragmentation during yielding remain scarce. Bulk rheology cannot discriminate between mechanisms---bond rupture, cluster rearrangements, or cooperative network reorganization---that produce similar macroscopic signatures. Three questions persist: What length and time scales govern irreversible rearrangements? How does the nature of structural relaxation evolve as networks yield? And can we identify a physical mechanism that quantitatively links nanoscale dynamics to macroscopic observables like $\tan \delta$ across the yielding transition?

%%%%%%%%%%%%%%%  state of the art echo-rheo-xpcs, dls, ddm
Here, we introduce rheo-echo X-ray photon correlation spectroscopy (rheo-echo-XPCS), synchronized with oscillatory rheology, to directly probe nanoscale irreversible dynamics during yielding of a colloidal fractal gel. While scattering and microscopy under \textit{in situ} shear provide insight into average structural changes~\cite{masschaele2011,hipp2021,bauland2024}, access to the dynamics of the microstructure during deformation is essential for understanding yielding and flow. Rheo-echo methods based on diffusing-wave spectroscopy~\cite{hebraud1997}, differential dynamic microscopy~\cite{aime2019,richards2021,edera2021}, dynamic light scattering~\cite{petekidis2002,laurati2014}, and XPCS~\cite{rogers2014,leheny2015,rogers2018,kamani2025} probe the temporal evolution of the microstructure via the intensity autocorrelation function $g_2(\Delta t,q)$.
Under oscillatory shear, deformation-induced particle displacements cause $g_2(\Delta t,q)$ to decay; for purely elastic and reversible deformation, the microstructure is recovered after each cycle, producing ``echo'' peaks at integer multiples of the oscillation period. Deviations from reversibility—arising from irreversible rearrangements or accumulated plasticity or flow from one strain oscillation to the next—lead to attenuation of the echo amplitude, providing a sensitive measure of microscopic irreversibility~\cite{aime2019,richards2021,edera2021,laurati2014,petekidis2002,rogers2014}.
To isolate irreversible dynamics, we exploit the vorticity direction, which is free of affine shear displacement in oscillatory Couette flow and immune to flow-reversal averaging inherent to double-pass rheo-XPCS geometries, in contrast to the flow and gradient directions where deterministic advection dominates~\cite{gadala1980}. 
Finally, to overcome problems of linear XPCS acquisition~\cite{rogers2014} such as the temporal-resolution and radiation-damage limitations 
or to perform experiments on long time scales without stitching data from separate acquisitions, we implement a nonlinear stroboscopic acquisition scheme enabling time-resolved echo measurements over extended durations (see SM.D~\cite{SMbib}).

%\section{Results}
%\vspace{2 mm}
%%%%%%%%%%%%%%%%%%%%%%%%%%%%%%%%%%%%

\section{Rheo-Echo-XPCS experiment on a CB gel}

We study colloidal gels formed by carbon black particles (CB, Vulcan\textsuperscript{\textregistered} PF) with a radius of gyration $r_0 \simeq 75~\mathrm{nm}$ (see SM.A~\cite{SMbib}). When dispersed in light mineral oil, their attractive interactions are dominated by van der Waals forces, with an estimated depth $U \simeq 25~k_B T$ and range $\delta \simeq 6.7~\mathrm{nm}$~\cite{dagastine2002,Trappe2007,bauland2025b}. At a concentration $c_w= 6~\%$, the particles form a yield-stress fluid—a space-spanning fractal gel~\cite{trappe2000,bauland2024,bauland2025} that behaves as a soft solid and can be rejuvenated by strong shear~\cite{dages2022,bauland2024}, similarly to depletion gels~\cite{Koumakis2015}.

%%%%%%%%%%%%%%%%%%%%%%%%%%%%%%%%%%%%
 \begin{figure}
    \centering
    \includegraphics[width=0.45\textwidth]{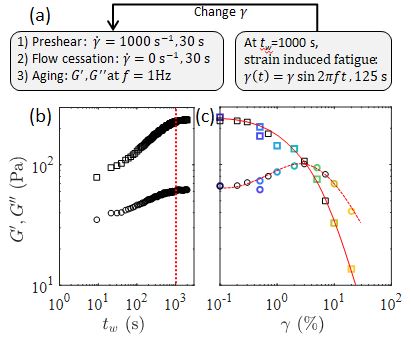}
    \caption{Rheology of a carbon black gel at $c_w=6$\%. (a) Protocol: the sample is presheared, then flow is stopped to allow gelation and aging. At $t_w=1000$~s, an oscillatory strain $\gamma(t) = \gamma \sin(2\pi f t)$ is applied with frequency $f = 1$~Hz and strain amplitude $\gamma$. The sequence is repeated for different $\gamma$. (b) Time-evolution of the of the elastic $G'$ ($\square$) and viscous $G''$ ($\circ$) moduli during the aging step. The dash vertical line corresponds to $t_w=1000$~s.  
(c) Strain dependence of the viscoelastic moduli during the strain-induced fatigue step ($t_w = 1000~\mathrm{s}$). The values of $G'$ and $G''$ are averaged over $t = 3$--$103~\mathrm{s}$, and each colored data point is associated with a simultaneous XPCS measurement (Fig.~\ref{fig2:xpcs}).
}
    \label{fig1:rheo}
\end{figure}
%%%%%%%%%%%%%%%%%%%%%%%%%%%%%%%%%%%%

The CB sample was loaded into a stress-controlled rheometer (Haake RS6000 Thermo Scientific) equipped with a polycarbonate Couette (gap 1~mm, inner diameter $20~\rm mm$).
%(inner diameter $20~\rm mm$, outer diameter $22~\rm mm$, gap $e=1$~mm and height $40~\rm mm$). 
The rheological protocol, illustrated in Fig.~\ref{fig1:rheo}a, consists of four steps.
(i) Preshear: the sample is presheared at a shear rate of $\dot\gamma = 1000$~s$^{-1}$ for 30~s to rejuvenate the microstructure. 
(ii) Flow cessation: shear is abruptly stopped by imposing $\dot\gamma = 0$~s$^{-1}$ for 30~s, allowing the CB dispersion to form a gel.
(iii) Aging monitoring: the time evolution of the gel is monitored by measuring the elastic $G'$ and the viscous $G''$ moduli, using small-amplitude oscillatory shear at $f = 1$~Hz and $\gamma = 0.1$~\%. 
Figure~\ref{fig1:rheo}b shows that $G'$ dominates $G''$ throughout the measurement, and both moduli increase with time, indicating progressive aging of the gel. Beyond a waiting time of $t_w = 1000$~s, the aging dynamics slows noticeably. This protocol based on continuous pre-shear rejuvenation is highly reproducible (see SM.B and~\cite{bauland2025b}) and serves our purpose of systematically characterizing irreversible dynamics under controlled driving conditions, though we acknowledge that alternative protocols for step (i) and (ii) may yield different gel microstructures with distinct yielding behaviors~\cite{edera2025,Koumakis2015}. This protocol results in CB gel with a multiscale structure: CB particles aggregate into primary fractal clusters, which further assemble into larger fractal clusters~\cite{Teixeira1988,Richards2017,koga2005,koga2008,bouthier2022b,bauland2025,bauland2025b} (see SM.C~\cite{SMbib}). 
These large clusters percolate to form a stress-bearing network throughout the material leading to a soft solid with viscoelastic properties.
(iv) Fatigue experiment: at $t_w = 1000$~s, a fatigue experiment is initiated by applying an oscillatory strain of amplitude $\gamma$ at $f = 1$~Hz for 125~s. During this period, the viscoelastic moduli remain stable (see SM.B~\cite{SMbib})
, and their time-averaged values are reported in Fig.~\ref{fig1:rheo}c. 
In Fig.~\ref{fig1:rheo}c, we display the viscoelastic moduli as a function of the fatigue strain amplitude $\gamma$, obtained by repeating the entire rheological protocol with values of $\gamma$ from $0.1$, to $20$~\%. This is the equivalent of a point by point strain sweep experiment. As the strain \tg{amplitude} increases, we observe that $G'$ slowly decreases then drastically descreases as the loss modulus overcome the elastic modulus as the \tg{conventional} yield strain $\gamma_y\simeq3.5$~\% defined when $G'(\gamma_y)=G''(\gamma_y)$. 

%%%%%%%%%%%%%%%%%%%%%%%%%%%%%%%%%%%%
 \begin{figure}
    \centering
    \includegraphics[width=0.43\textwidth]{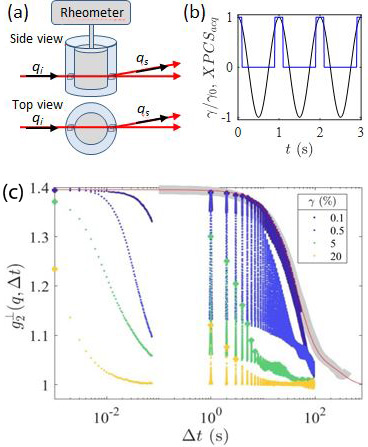}
    \caption{Rheo-Echo-XPCS. (a) The incident X-ray beam (red arrows) go through the center of the couette cell. The scattered intensity is collected on a 2D detector. (b) Rheology (black line) and XPCS acquisitions (blue line) are synchronized so that XPCS acquisition are carried out periodically in bunch mode of duration 0.2~s and of frequency 1~kHz. The bunches are centered on the max strain amplitude. We use a hardware trigger between the rheometer and the beamline electronics to synchronize the acquisition. (c) Evolution of the intensity autocorrelation function in the vorticity direction $g_2^{\perp}(q,\Delta t)$ with the lag time $\Delta t$ for strain amplitudes $\gamma=0.1$, 0.5, 5, 20~\% at $q=0.016$~nm$^{-1}$. The + symbol correspond to maxima of the echo peak. The grey line corresponds to a classical XPCS acquisition (5000 images at 10~Hz) performed on the CB gel at rest in the Couette geometry and averaged over all $q$ directions.  }
    \label{fig2:xpcs}
\end{figure}
%%%%%%%%%%%%%%%%%%%%%%%%%%%%%%%%%%%%

%%%%%%%%%%%%%%%%%%%% g2echo

At the start of the fatigue step, we initiate a 100~s X-ray Photon Correlation Spectroscopy experiments on beamline ID02 (ESRF, Grenoble, Fr)~\cite{narayanan2022,chevremont2024}. 
%The incident X-ray beam, with a wavelength of approximately $0.1$nm ($E = 12.23$keV), was collimated to a beam size of $25\mu$m (vertical) by $20\mu$m (horizontal).
Two-dimensional scattering patterns, $I(q, t)$, were recorded using an Eiger2 4M pixel array detector, where $q$ denotes the scattering vector magnitude and $t$ the acquisition time.
As sketched in Fig.~\ref{fig2:xpcs}a, we used a radial scattering configuration: the X-ray beam was aligned to intersect the rotational axis of the rheometer, defining two privileged directions in the scattering plane: the flow direction ($e_{\parallel}$) and the vorticity direction ($e_{\perp}$).

XPCS measurements are performed using a new  acquisition scheme. Rheology and XPCS are stroboscopically synchronized. XPCS data are collected periodically in bunch mode: short acquisitions of duration 0.2s were performed at 1~kHz, as shown in Fig.\ref{fig2:xpcs}b. 
Each bunch is centered on the time of maximum strain amplitude, i.e., at \( t = n T \), with \( n \in \mathbb{N} \) and $T=1/f=1~s$.
From the full set of 20,000 frames of $I(\vec{q}, t)$, we compute the two-time intensity autocorrelation function $g_2(q, \Delta t, t)$ in the vorticity direction, where $\Delta t$ is the lag time and $t$ the experimental time (see SM.D~\cite{SMbib}). \tg{As shown in SM.D~\cite{SMbib}, $g_2(q, \Delta t, t)$ exhibits stationary statistics throughout the measurement window, justifying time-averaging over $t$}. We therefore define the vorticity-direction autocorrelation function as $g_2^{\perp}(q, \Delta t) = \frac{\left\langle I(\vec{q}, t) \, I(\vec{q}, t + \Delta t) \right\rangle}{\left\langle I(\vec{q}) \right\rangle^2},$ where $\langle \cdot \rangle$ denotes the ensemble average over all detector pixels with wavevectors $\vec{q}$ oriented within $\pm 10^\circ$ of the vorticity direction and within a specified range of $|\vec{q}|$ and average over the time $t$. The intermediate scattering function $g_1^{\perp}(q, \Delta t)$ is given by the Siegert relation, $g_2^{\perp}(q, \Delta t)=1+\beta g_1^{\perp}(q, \Delta t)^2$ where $\beta$ is the speckle contrast. 
%%%%%%%%%%%%%%%%%%%%%%%%%%%%%%%%%%%%
 \begin{figure*}
    \centering
    \includegraphics[width=0.9\textwidth]{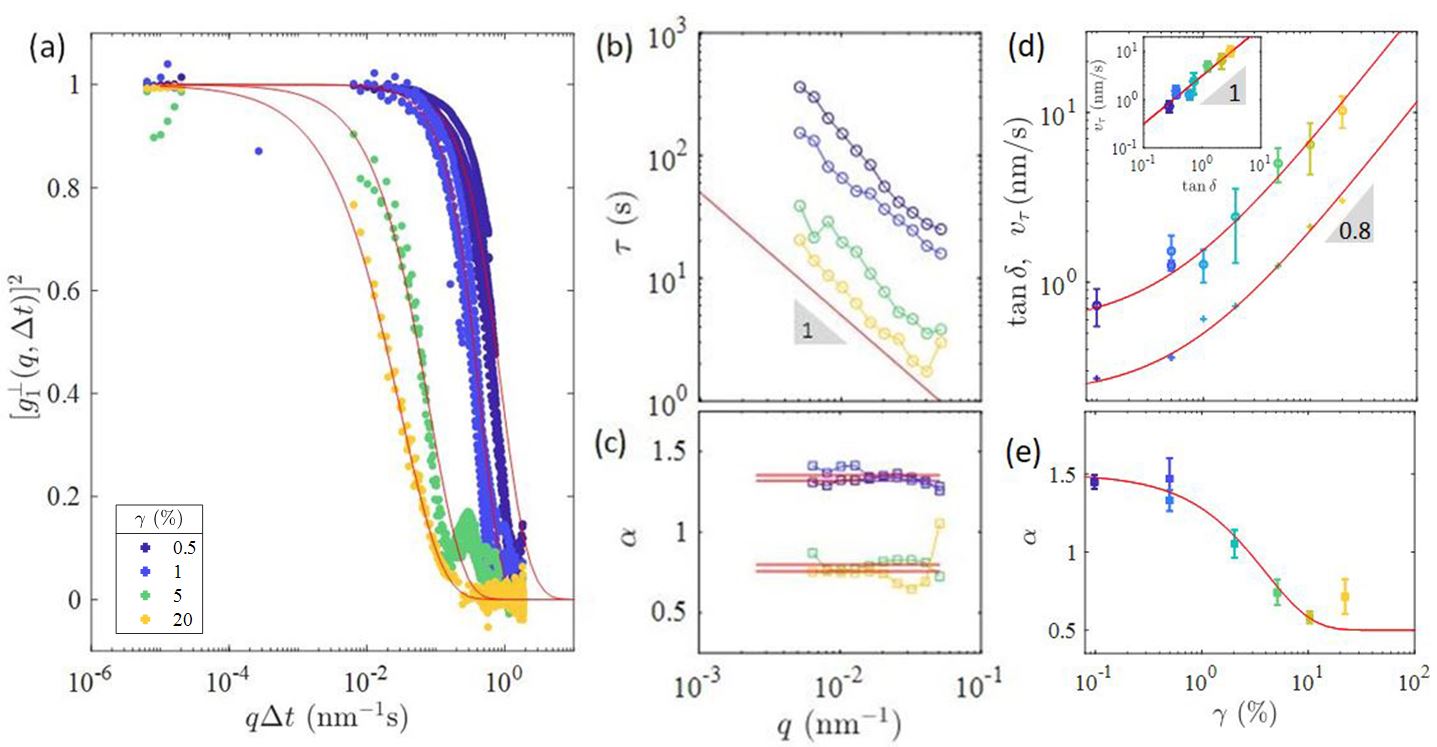}
    \caption{(a) Intensity autocorrelation function in the vorticity direction $g_2^{\perp}(q,\Delta t)$ taken at the maximum of the echo peak at $\Delta t=nT$ (extracted from fig.~\ref{fig2:xpcs}) as function of $q\Delta t$ for 5 different $q$ ranging from $0.006$ to  0.051~nm$^{-1}$. Red lines are averaged fit over $q$ to data using  $g_1^{\perp}(q,\Delta t)=e^{-(\Delta t/\tau)^\alpha}$ with $\tau(q)$ the decoraletion time and $\alpha(q)$ its associated exponent. (b) $q-$dependency of $\tau$. (c) $q-$dependency of $\alpha$. (d) Evolution of $\tan\delta=G''/G'$ (+) based on the measurements in Fig.~\ref{fig2:xpcs}(c) and evolution of the $q$-averaged value of the decorrelation velocity $v_{\tau}=1/\langle  q \tau\rangle$ ($\circ$) with $\gamma$. Inset: $v_{\tau}$ versus $\tan\delta$. The red line is a linear fit to the data: $v_{\tau} = v_0\tan\delta$, with $v_0=3.1\pm 0.2$ nm/s. (d) Evolution of the $q$-averaged value of $\alpha$ as function of the strain. In (d-e) Red lines are a guide to the eyes. Error bars are the standard deviation resulting from data collected at different $q$.}
    \label{fig4:xpcs}
\end{figure*}
%%%%%%%%%%%%%%%%%%%%%%%%%%%%%%%%%%%%
%\section{Results}

\vspace{2 mm}
%%%%%%%%%%%%%%% echo peak amplitude
\section{Evolution of the echo autocorrelation through the yielding transition}
%{\textit{Yielding transition -- }}
The evolution of $g_2^{\perp}(q, \Delta t)$ as function of $\Delta t$ is shown in Fig.~\ref{fig2:xpcs}c. As expected, the intensity autocorrelation function exhibits distinct echo peaks, regularly spaced by the oscillatory strain period of 1~s. 
%This modulated structure in $g_2^{\perp}$ arises from the periodic recovery of the material's microscopic configuration: at each maximum strain, the system transiently returns to a similar we show the evolution, resulting in enhanced correlation at $\Delta t = 1, 2, \ldots$~s. 
%echoes are due to the transit term + deformation of structure
Rheo-echo correlations provide two complementary measures of microscopic dynamics: the intra-cycle shape of individual echo peaks (see SM.D~\cite{SMbib}) and the cycle-to-cycle decay of their amplitudes (Fig.~\ref{fig4:xpcs}a). The temporal shape of an echo peak near $\Delta t = nT$ reflects the distribution of irreversible displacements accumulated within a single oscillation cycle. In contrast, the echo amplitude at $\Delta t = nT$ quantifies the cumulative loss of structural memory over many cycles, arising from the progressive accumulation of irreversible rearrangements that prevent full recovery and which is our focus. 
In Fig.~\ref{fig4:xpcs}a, we show the evolution $[g_1^{\perp}(q,\Delta t =nT)]^2$ at the maximum amplitude of the echo peaks as function of $q\Delta t$ for different strain amplitude $\gamma$ and $q-$amplitude. We observe that at a given $\gamma$,  $[g_1^{\perp}]^2$ at different $q$ ranging from 0.006 to 0.051~nm$^{-1}$ scale on a master curve indicating that the characteristic decorrelation time $\tau(q)$ is $q-$dependant and has a ballistic behavior. We note that the decay of $[g_1^{\perp}]^2$ becomes smoother (less abrupt) as $\gamma$ increases.
%\cite{SMbib}).
Focusing on the short time scales, we fit $[g_1^{\perp}]^2$ from $[g_1^{\perp}]^2=1$ to 0.4 using a compressed/stretched exponential: $g_2^{\perp}(q,\Delta t)
=1+\beta\bigl[g_1^{\perp}(q,\Delta t)\bigr]^2$, with $g_1^{\perp}(q,\Delta t)=e^{-(\Delta t/\tau)^{\alpha}}$.
$\tau$ denotes the relaxation time, $\alpha$ is the exponent associated with it, and the average relaxation time is 
$\langle \tau \rangle = \frac{\tau}{\alpha} \, \Gamma\!\left(\frac{1}{\alpha}\right)$.
In Fig.~\ref{fig4:xpcs}b, we confirm quantitiatively the presence of ballistic scaling both at rest and during the fatigue experiments, with $\langle\tau\rangle \propto q^{-1}$. This defines a characteristic velocity $v_\tau = 1/\langle q\tau\rangle$ that sets the rate at which irreversible rearrangements decorrelate the gel dynamics.

In colloidal gels at rest, such ballistic behavior of the intermediate scattering function is commonly attributed to the rupture of inter-cluster bonds~\cite{cipelletti2000,cipelletti2003,gado2007}. These rupture events induce localized stress release, which propagates through the gel as elastic waves, resulting in microscopic displacements and a rapid loss of correlation in the scattering signal. During fatigue, this ballistic scaling of $\tau$ persists, indicating that bond snapping between clusters continues to be a dominant mechanism of decorrelation under oscillatory strain in this $q$-range. However, we observe that $\tau$ systematically decreases as the strain amplitude $\gamma$ increases, suggesting that deformation accelerates the rupture of inter-cluster bonds and enhances the rate of structural decorrelation.
This decrease in $\tau$ with increasing $\gamma$ is accompanied by a reduction in $\alpha$ (Fig.~\ref{fig4:xpcs}(c, e)), shifting the dynamics from a compressed exponential ($\alpha \simeq 1.5$) at low $\gamma$ to a stretched exponential ($\alpha \simeq 0.5$) at high $\gamma$. 
For colloidal gels in quiescent conditions, Cipelletti et al.~\cite{cipelletti2000} observed a ballistic relaxation with $\alpha = 1.5$ and rule out the bond-breaking mechanism, which typically rely on $\alpha = 2$. They propose that this dynamics arises from syneresis-driven restructuring, where $\alpha=1.5$ emerges from three-dimensional dipolar strain fields created by localized shrinkage inhomogeneities. These inhomogeneities produce displacement fields that fall off as inverse distance squared and lead to volume-scaling decorrelation probabilities around each defect.
We extend the model of~\cite{cipelletti2000} by introducing an effective $dynamic$ decorrelation dimensionality $d_f$ for the number of scatterers affected by rearrangements (see SM.D~\cite{SMbib}).
%SM~\cite{SMbib}).
This modification leads to a self-consistent intermediate scattering function with ballistic decorrelation times ($\tau\propto q^{-1}$) and an exponential exponent $\alpha=d_f/2$. Within this framework, yielding is interpreted as a progressive fragmentation of the stress-bearing network such that irreversible rearrangements contributing to decorrelation along the vorticity direction are transmitted through an increasingly sparse and filamentary backbone. This phenomenological picture is captured by a reduction of the effective vorticity-projected decorrelation dimension from $d_f \simeq 3$ to $d_f \simeq 1$, rationalizing the crossover from compressed ($\alpha \simeq 1.5$) to stretched ($\alpha \simeq 0.5$) relaxation while preserving ballistic-like scaling of the relaxation time. Importantly, $d_f$ is a dynamical effective dimension associated with stress transmission and decorrelation in the vorticity direction, rather than a static structural fractal dimension. While such filamentary pathways may not be directly visible in the static structure factor, simulations of colloidal gels~\cite{colombo2014} show that stress localizes into sparse load-bearing structures under deformation, supporting this interpretation.

The evolutions of the correlation-function shape is not unique to the present system and has been reported in several classes of dynamically arrested materials. However, the associated wave-vector dependence of the relaxation time differs substantially between systems, suggesting that the evolution of the correlation-function shape may arise from distinct microscopic mechanisms. Simulations varying the thermal-to-bond energy ratio ($k_B T/\varepsilon$) reported a transition from compressed to stretched dynamics accompanied by diffusive relaxation ($\tau\propto q^{-2}$)~\cite{bouzid2017}.
%%%
\tg{Rogers \textit{et al.}~\cite{rogers2018} also reported strain-dependent changes in the line shape of the echo correlation function in concentrated nanoemulsions under oscillatory shear. In their system, the relaxation was generally described by either simple or compressed exponentials ($\alpha \simeq 1$--1.7), depending on droplet concentration and strain amplitude, together with a sub-ballistic wave-vector dependence $\tau(q)\sim q^{-0.7}$. 
We attribute these differences to the distinct nature of stress transmission in the two dynamically arrested systems. Rogers \textit{et al.} studied a densely packed emulsion with glass-like stucture, whereas the present system consists of a fractal gel network.}

%%%%%%%%%%%%%%%%%%%%%%%%%%%%%%
\section{Quantitative relationship between the rheology and XPCS}

Finally, we explore the quantitative relationship between the rheology and XPCS observables. In Fig.~\ref{fig4:xpcs}d, we compare the evolution of $\tan\delta = G''/G'$ with the decorrelation velocity $v_{\tau} = 1/\langle q\tau \rangle$ as functions of strain amplitude $\gamma$, measured simultaneously during the same fatigue experiment. As $\gamma$ increases, $\tan\delta$ first displays a plateau in the linear viscoelastic regime and subsequently increases as a power law. Remarkably, within error bars, $v_{\tau}$ follows the same evolution quantitatively, yielding $v_{\tau} = v_0\tan\delta$ with $v_0=3.1$ nm/s, throughout the full range explored (Fig.~\ref{fig4:xpcs}d, inset).
\tg{The ballistic-like dynamics should not be interpreted as a literal particle transport velocity or as evidence for persistent particle trajectories. Rather, $v_{\tau}$ represents an effective microscopic configurational renewal rate associated with the irreversible loss of structural memory under cyclic loading, consistent with the irreversible strain framework developed in Sec.~D of~\cite{SMbib}. Remarkably, the proportionality $v_{\tau}\propto\tan\delta$ persists across both linear and nonlinear regimes despite the substantial evolution of the microscopic dynamics and the apparent crossover in the effective dimensionality of the stress-bearing network.}

\tg{This persistent proportionality deserves careful interpretation. In principle, $\tan\delta$ reflects dissipation arising from multiple microscopic processes, including viscoelastic relaxation within the transient network and irreversible rearrangements associated with yielding and progressive damage under cyclic loading~\cite{donley2020,kamani2021}. If $v_{\tau}$ were sensitive only to large-scale plastic yielding events, one might expect the proportionality between $v_{\tau}$ and $\tan\delta$ to break down in the nominally linear viscoelastic regime. Instead, the continuity of the proportionality suggests that slow irreversible structural fluctuations and progressive configurational renewal already contribute to both the finite dissipation and the nonzero echo decorrelation measured experimentally, even below the conventional yielding crossover at $\gamma_y$.}

\tg{These results provide a microscopic perspective complementary to recent continuum and intra-cycle descriptions of yielding under oscillatory shear~\cite{donley2020,kamani2021}. Kamani, Donley and Rogers~\cite{kamani2021} introduced a viscoplastic framework based on the decomposition of deformation into recoverable and unrecoverable strain components within oscillation cycles, allowing plastic deformation and smooth $G''$ overshoots to emerge continuously below the nominal yield stress. More recently, Keane \textit{et al.}~\cite{keane2025} demonstrated experimentally that key nonlinear yielding signatures can be predicted directly from linear viscoelastic properties such as $\tan\delta$. Our observation that $v_{\tau}\propto\tan\delta$ across both the linear and nonlinear regimes provides a microscopic structural counterpart to these macroscopic findings, suggesting that macroscopic dissipation remains continuously linked to irreversible configurational renewal even as the network geometry and microscopic relaxation mechanisms evolve substantially under shear.}

\tg{Our rheo-echo-XPCS measurements are conceptually distinct from recent time-resolved rheo-XPCS approaches~\cite{kamani2025}. In~\cite{kamani2025}, the two-time correlation function $g_2(q,\Delta t,t)$ is evaluated within individual oscillation cycles and integrated over scattering directions under oscillatory shear. This intra-cycle approach probes transient structural reversibility during a single deformation period and relates microscopic structural memory to macroscopic rheological response through the balance between recoverable and unrecoverable deformation within a cycle.
In contrast, the rheo-echo protocol used here probes correlations between successive oscillation cycles. By measuring echoes at fixed macroscopic strain in the vorticity direction, reversible affine contributions are strongly suppressed, such that the measured decorrelation directly reflects cycle-to-cycle configurational renewal.
These two approaches therefore probe fundamentally different temporal hierarchies of the dynamics. Time-resolved rheo-XPCS emphasizes structural dynamics within a single oscillation cycle and assumes that each cycle probes a slowly evolving structural state. In this framework, reversible and irreversible processes coexist during a cycle, giving rise to spatially heterogeneous reversible dynamics in which only subsets of the microstructure contribute to recoverable deformation. By contrast, rheo-echo-XPCS directly accesses the accumulation of microscopic irreversibility across cycles, i.e., the fatigue-induced evolution of the structural state dynamics itself.
This distinction is particularly relevant for fatigue-like processes, where yielding is governed not only by reversible structural fluctuations within a cycle, but also by the progressive buildup of irreversible configurational changes from cycle to cycle, leading to structural memory loss and network reorganization under repeated shear.}

%%%%%%%%%%%%%%
Finally, the dimensional crossover $d_f: 3 \rightarrow 1$ suggests that successful models of gel fatigue must account not only for the rate of damage accumulation---captured macroscopically by $\tan\delta$---but also for the evolving geometry of stress transmission as the network progressively reorganizes under shear. The persistence of the $v_{\tau} \propto \tan\delta$ proportionality despite this crossover indicates that macroscopic dissipation remains tightly coupled to the microscopic irreversibility rate even as the spatial organization of the dynamics evolves from three-dimensional collective rearrangements toward more filamentary stress-bearing pathways. Combining rheo-echo-XPCS with complementary structural probes such as confocal microscopy or particle-resolved simulations could help determine whether the observed dimensional reduction reflects genuine filament formation, heterogeneous stress localization, or effective coarse-graining of stress transmission in progressively damaged networks.

%%%%%%%%%%%%%%%%%%%%%%%%%%%

%\vspace{2 mm}
%\textit{Conclusion} -- 
\section{Conclusion}

\tg{We investigate the mechanical response of a colloidal gel under oscillatory shear at constant strain amplitude $\gamma$, and show that rheo-echo-XPCS simultaneously and quantitatively resolves the cycle-to-cycle accumulation of nanoscale irreversible configurational renewal.}

\tg{From the XPCS echo measurements performed in the vorticity direction, the autocorrelation function exhibits a ballistic-like decay of the echo peak carateristic time $\tau$ at all strain amplitude $\gamma$, allowing us to define a microscopic decorrelation rate $v_\tau(\gamma)$. 
$v_\tau$ directly quantifies the rate of irreversible structural rearrangements. As the strain amplitude increases, the temporal decay of the echo-autocorrelation function evolves from compressed to stretched exponential behavior ($\alpha: 1.5 \rightarrow 0.5$). 
Within a framework in which density correlations are governed by localized rearrangements acting as elastic dipoles, this crossover is interpreted as a progressive reduction of the effective dimensionality of the stress-bearing network ($d_f: 3 \rightarrow 1$).}

\tg{Comparing rheological and microscopic measurements, we find that the macroscopic loss tangent $\tan\delta$ is directly proportional to $v_\tau$, a relation that remains valid from the linear viscoelastic regime to strongly nonlinear deformation and beyond $\gamma_y$. In this view, the loss tangent $\tan\delta$ provides a macroscopic metric directly proportional to the rate of irreversible configurational renewal of the microstructure, manifesting as a progressive loss of structural memory under cyclic loading.}

More broadly, rheo-echo-XPCS provides a new route to investigate how network geometry, structural memory, and irreversible dynamics couple during fatigue and yielding in soft arrested materials.

%We demonstrate that the macroscopic loss tangent ($\tan \delta$)  quantitatively tracks the cycle-to-cycle accumulation of nanoscale irreversible configurational renewal measured} by rheo-echo-XPCS, with the proportionality $v_\tau \propto \tan \delta$ remaining valid from the linear viscoelastic regime to strongly nonlinear deformation as the strain amplitude increases.
%As the strain amplitude increases, the structural relaxation evolves from compressed to stretched exponential behavior ($\alpha: 1.5 \rightarrow 0.5$) while maintaining a ballistic-like scaling of the decorrelation time. Within a framework where the field correlations are governed by localized rearrangements acting as elastic dipoles, this evolution is interpreted as a progressive reduction of the effective dimensionality of the stress-bearing network ($d_f: 3 \rightarrow 1$). 
%\tg{Our results establish a direct quantitative connection between macroscopic dissipation and the cycle-to-cycle accumulation of microscopic irreversibility under oscillatory shear. More broadly, rheo-echo-XPCS provides a new route to investigate how network geometry, structural memory, and irreversible dynamics couple during fatigue and yielding in soft arrested materials.}

\section*{authors contribution}
All the authors participated to the rheo-XPCS experiments and the discussion of the data. WC set up the rheo-echo-XPCS experiments, the bunch mode acquisition and the calculation of the intensity auto correlation function. TG designed the project, analyzed the data and wrote the article. 

\section*{Acknowledgements}
The authors are especially grateful to the ESRF for beamtime at the beamline ID02 (proposal SC-5524) and Theyencheri Narayanan for the discussions. 
This work was supported by the grants: ANR-21-CE06-0020-01 and FWF/ANR-TrainGel as well as the European Union’s Horizon Europe Framework Program HORIZON under the Marie Skłodowska-Curie Grant Agreement 101120301 and the LabEx iMUST of the University of Lyon (ANR-10-LABX-0064). This work benefited from meetings within the French GDR CNRS 2019 `Solliciter LA Matière Molle' (SLAMM). 

%\bibliographystyle{apsrev4-1}
%\bibliography{biblio}

%\begin{comment}

%%%%%%%%%%%%%%%%%%%%%%%%%%%%%%%%%%%%

\newpage

\section*{Supplemental Material (SM), Tracking microscopic irreversibility during yielding of a colloidal fractal gel with Rheo-Echo-XPCS}

\renewcommand{\thefigure}{S\arabic{figure}}
\setcounter{figure}{0}

The supplemental material is composed of four sections. 
Section A describes the preparation and physical properties of the carbon black (CB) particles used to form the gel, including their size, fractal structure, and volume fraction.
Section B explains the rheological protocol, emphasizing the reproducibility of the preshear--aging procedure and the rationale for using a fatigue-based oscillatory method instead of a standard strain sweep. 
Section C presents the static structural characterization of the gel using small-angle X-ray scattering (SAXS), modeled with a hierarchical fractal framework.
Section D details the rheo-echo-XPCS methodology, including the focus on the vorticity direction to isolate non-affine irreversible dynamics and the nonlinear acquisition scheme synchronized with oscillatory shear emphasizing its originality with respect to the literature. It then presents the two time autocorrelation function with emphasis on the stationary of the experiment. Finally, it introduces the model linking ballistic decorrelation and exponent $\alpha$, with the decorelation velocity $v_{\tau}$ and the effective stress-bearing dimensionality $d_f$ during yielding.

{\subsection{Carbon black particle}} 
Following~\cite{dages2022, bauland2024, bauland2025}, carbon black (CB) particles (Vulcan\textsuperscript{\textregistered} PF, Cabot; density $d_{cb} = 2.26 \pm 0.03$) were dispersed in mineral oil (RTM17 Mineral Oil Rotational Viscometer Standard, Paragon Scientific; viscosity $\eta_f = 252.1$ mPa$\cdot$s at $T = 25^{\circ}$C; density $d_{oil} = 0.871$) at a mass fraction $c_w = 6~\% $. The corresponding volume fraction of CB particles, $\phi$, was calculated as $\phi = c_w/[c_w + (d_{cb}/d_{oil})(1 - c_w)]\simeq0.024$
After mixing, the dispersions were sonicated for 2 hours in an ultrasonic bath (Ultrasonic Cleaner, DK Sonic\textsuperscript{\textregistered}, United Kingdom) to ensure complete dispersion of the particles.
Vulcan\textsuperscript{\textregistered}PF CB particle are composed are composed of nodules of radius $a\simeq 2$~nm that are fused together to form an aggregate of radius $r_0 \simeq 75$~nm with a fractal dimension $d_{f_0}\simeq 2.9$.
\vspace{2mm}

{\subsection{Rheology}} 

\begin{figure}[h]
    \centering
    \includegraphics[width=0.35\textwidth]{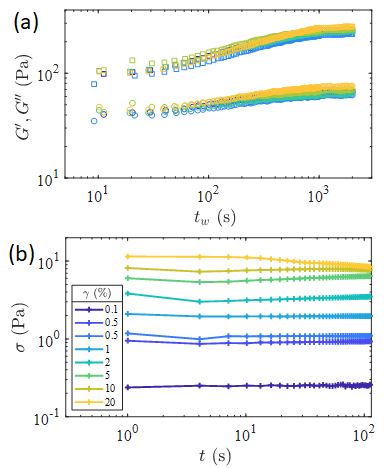}
    \caption{(a) Evolution of $G'$ ($\square$) and $G''$ ($\circ$) during the aging step. Colors code for eight iteration of the rheology protocol as described in Fig~.1a. (b) Time evolution of the stress amplitude $\sigma$ during the fatigue step as described in Fig~.1a.}
    \label{figS1:sigma}
\end{figure}

While different rejuvenation protocols---whether based on sinusoidal or continuous preshear---yield quantitatively different initial microstructures~\cite{bauland2025b, edera2025, Koumakis2015}, the key requirement for our comparative study is consistency and reproducibility, which our protocol reliably provides. As shown in Fig.~\ref{figS1:sigma}(a), the aging behavior of the viscoelastic moduli is consistently recovered across repeated runs, and the stress response remains stable throughout the XPCS measurement window ($t = 1$ to 100~s) for all strain amplitudes except at the largest strain $\gamma = 20\%$, where stress begins to decrease after $t \approx 20$~s (Fig.~\ref{figS1:sigma}(b)).
%However, at this strain amplitude, the structural decorrelation is already complete within approximately 20~s and therefore does not affect the measurement of the intensity autocorrelation function.
We employ a fatigue-based protocol rather than a conventional amplitude sweep for several reasons. First, standard strain sweeps combine two effects simultaneously: increasing strain amplitude \emph{and} accumulating damage over time, making it difficult to disentangle instantaneous nonlinear response from history-dependent structural evolution. Our protocol---applying constant-amplitude oscillations at fixed frequency for extended duration (125~s)---isolates the temporal accumulation of irreversibility at each strain level, enabling direct measurement of structural memory loss through echo decay. Second, this approach is essential for rheo-echo-XPCS: echoes require periodic driving at fixed amplitude to establish a reference state for reversibility, which is incompatible with continuously varying strain. Third, cyclic loading at fixed amplitude is representative of many real-world applications involving repeated mechanical stress. The resulting point-by-point characterization is equivalent to a strain sweep experiment but with superior control over aging time and structural history at each measurement point, allowing us to systematically isolate the effects of driving conditions on irreversible dynamics.

%%%%%%%%%%%%%%%%%%%%%%%%%%%%%%%%%%%%
%%%%%%%%%%%%%%%%%%%%%%%%%%%%%%%%%%%%

\begin{figure}[h]
    \centering
    \includegraphics[width=0.45\textwidth]{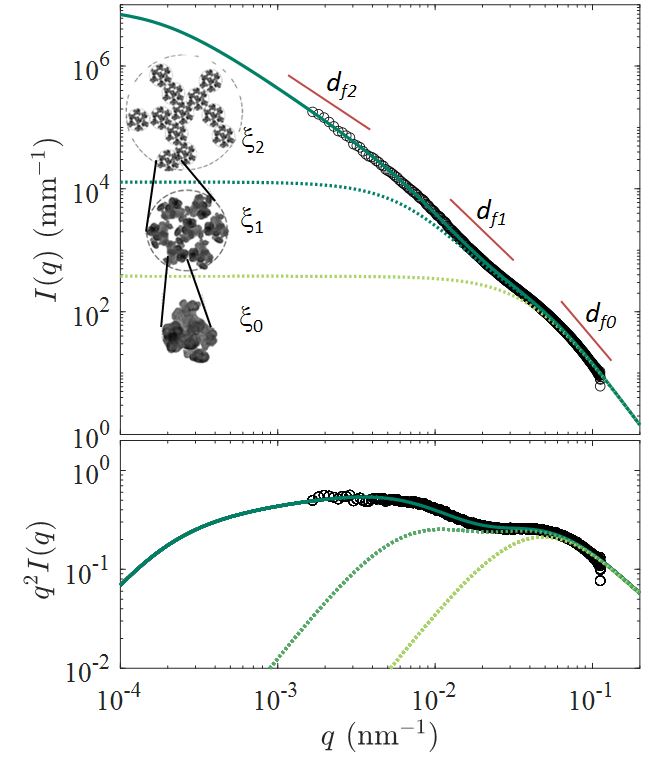}
    \caption{ CB gel structure at rest. Top: Averaged scattering intensity $I(q)$ vs wave vector $q$,   by the CB gel at $t_w=1000$~s just before initiating the fatigue experiment. Bottom: Kratky representation. The light green dash line is the form factor ($I\sim PS_0$) of CB particle of size $\xi_0$, the green dash line is the first level of the fit corresponding to cluster of size $\xi_1$ ($I\sim PS_0S_1$). The dark green line is the full fit ($I\sim PS_0S_1S_2$). }
    \label{figS2:saxs}
\end{figure}

\vspace{2mm}
{\subsection{SAXS}} 
In Fig.~\ref{figS2:saxs} we show the static scattering intensity as function of $q$ measured at $t_w=1000$~s just before performing the fatigue experiment. The structure of CB dispersions, as probed by USAXS, is modeled using a hierarchical framework described in~\cite{bouthier2022b, bauland2025}, which accounts for the organization of the CB particles (size $xi_0$, fractal dimension $d_{f_0}$) into small fractal clusters (size $\xi_1$, fractal dimension $d_{f_1}$), which further aggregate into a larger-scale fractal network (size $\xi_2$, fractal dimension $d_{f_2}$). Each level is described by a mass fractal structure factor $S_i$~\cite{Teixeira1988}. As previously reported, the CB particles themselves are composed of small ``nodules'' (size $a$) fused together~\cite{Richards2017,hipp2021}. The form factor of the CB particle is thus described as the product of the form factor of a sphere $P(q)$ and a mass fractal structure factor $S_0(q)$. The full model reads:
\begin{equation}
    I\left(q\right)= \phi_a V_a(\Delta\rho)^2 
    \underbrace{P(q) 
    S_0(q)}_{\text{$\xi_0$}} \underbrace{S_1(q)}_{\text{$\xi_1$}} \underbrace{S_2(q)}_{\text{$\xi_2$}}
\end{equation}

with
\begin{equation}
\begin{array}{l}
\left\{
\begin{array}{ll}
\text{ } & a < \xi_0 < \xi_1 < \xi_2 \\
P(q) & = \bigg[\frac{3[\sin(qa) - qa\cos(qa)]}{(qa)^3} \bigg]^2 \\
S_i(q) & = 1 + \frac{d_{f_i} \Gamma(d_{f_i} - 1)}{[1 + 1/(q\xi_i)^2]^{(d_{f_i} - 1)/2}}  \frac{\sin[(d_{f_i} - 1)\tan^{-1}(q\xi_i)]}{(q R_i)^{d_{f_i}}}
\end{array}
\right.
\end{array}.
\label{eq:SAXS}
\end{equation}

\noindent where $\phi_a$ and $V_a$ denote the volume fraction and unit volume of the nodules of size $a$, respectively, and $\Delta \rho$ is the scattering length density difference between mineral oil and CB particles. Approximating the CB particle as rough sphere, we have $\phi_a \approx \phi_{\xi_0}$.
$R_i$ represents a cutoff length, smaller than the corresponding cluster size $\xi_i$ but larger than the maximum size captured in $S_{i-1}(q)$. We took $R_i=2\xi_{i-1}$.

%%%%%%%%%%%%%%%%%%%%%%%%%%%%%%%%%%%%
%%%%%%%%%%%%%%%%%%%%%%%%%%%%%%%%%%%%

We note that $\xi_2$ lies beyond the accessible $q$-range of the USAXS measurements and would require probing at much lower $q$ values to be directly resolved. However, $\xi_2$ can be estimated based on the number density of particles, $\rho = \frac{\phi}{V}$, where $V$ is the volume of a CB particle of radius $\xi_0$. Assuming that the large clusters densely and homogeneously fill the available space and evaluating the particle concentration within each cluster, $\rho$ can also be expressed as

\begin{equation}
\rho = \frac{\phi}{V} = \frac{\left(\frac{\xi_2}{\xi_1}\right)^{d_{f2}} \left(\frac{\xi_1}{\xi_0}\right)^{d_{f1}}}{\xi_2^3}.
\end{equation}

\noindent By equating these two expressions of $\rho$, $\xi_2$ can be expressed as a function of $\rho$, $\xi_0$, $\xi_1$, $d_{f_1}$, and $d_{f_2}$, and this expression can then be incorporated into the fit (eq.~\ref{eq:SAXS}).
For the present CB gel at $c_w=6$~\%, we find, $\xi_0=32$~nm, $d_{f_{0}}=2.9$, $\xi_1 =140$~nm, $d_{f_1}=2.6$, $\xi_2 =5600$~nm and $d_{f_2}=1.8$ (see Fig.~\ref{figS2:saxs}). The radius of gyration is given by $r_i=\sqrt{d_{f_i}(1+d_{f_i})/2}\xi_i$.

\vspace{2mm}
{\subsection{XPCS}}

\noindent\textbf{Vorticity direction } -- In oscillatory shear, dominant affine and reversible particle displacements occur in the flow and velocity-gradient directions. Along these directions, the scattering signal is strongly affected by deterministic shear advection and flow reversal, which leads to rapid decorrelation of the speckle pattern even in the absence of irreversible rearrangements. Although echo techniques partially compensate for affine motion, residual shear-induced phase shifts and small deviations from perfect reversibility make it difficult to disentangle true plastic events from reversible kinematics in these directions.
In contrast, the vorticity direction is, by symmetry, free of affine shear displacement. 
Indeed, for oscillatory Couette flow between concentric cylinders, the velocity field in cylindrical coordinates $(r, \theta, z)$ is $\mathbf{v}(\mathbf{r}, t) = v_\theta(r, t) \hat{\boldsymbol{\theta}}$, where the azimuthal velocity varies with radius. The velocity gradient tensor in the narrow-gap approximation (where the gap width $d \ll R$, the mean radius) reduces to the same form as planar shear: $\nabla \mathbf{v} \approx \begin{pmatrix} 0 & \dot{\gamma}(t) & 0 \\ 0 & 0 & 0 \\ 0 & 0 & 0 \end{pmatrix}$ in the local Cartesian frame $(x, y, z)$ corresponding to (azimuthal, radial, axial) directions, with $\dot{\gamma}(t) = r \, d\omega/dr \approx \dot{\gamma}_0 \cos(\omega t)$. The affine displacement rates are $dx_{\text{affine}}/dt = \dot{\gamma}(t) y$, $dy_{\text{affine}}/dt = 0$, and $dz_{\text{affine}}/dt = 0$. The axial direction $\hat{\mathbf{z}}$ serves as the vorticity direction, and the affine displacement along it vanishes: $\Delta z_{\text{affine}} = 0$ for all times and positions. By cylindrical symmetry, the azimuthal flow produces no axial advection, so any displacement observed along $\hat{\mathbf{z}}$ must arise from non-affine, irreversible rearrangements. This makes the axial (vorticity) direction the optimal probe of microscopic plasticity in Couette rheo-XPCS experiments, free from the confounding effects of deterministic shear advection that dominate the azimuthal and radial directions.

An other advantage of focusing on the vorticity direction is that it unsentitive to flow-reversal. Indeed, in the Couette geometry of Fig. 2a, the X-ray beam passes through the sample twice, causing the detector to record a superposition of scattering signals from regions experiencing opposite flow directions. This is problematic because many complex fluids exhibit flow-reversal anisotropy~\cite{gadala1980}, where the microscopic structure and dynamics differ between forward and reverse shear due to non-equilibrium distortions, shear-induced alignment, or history-dependent plastic rearrangements.
The measured signal along the flow direction therefore becomes a mixed average that does not cleanly represent either flow state, complicating quantitative interpretation of correlation functions and separation of reversible versus irreversible dynamics. 
However this ambiguity does not affect the vorticity direction because it is orthogonal to the flow velocity; the double-pass geometry probes identical structural/dynamical features regardless of flow direction along this axis. This makes the vorticity direction not only free from affine displacement but also free from geometric artifacts introduced by the dual-pass measurement, reinforcing its status as the most reliable probe of microscopic irreversibility.

\begin{figure}
    \centering
    \includegraphics[width=0.49\textwidth]{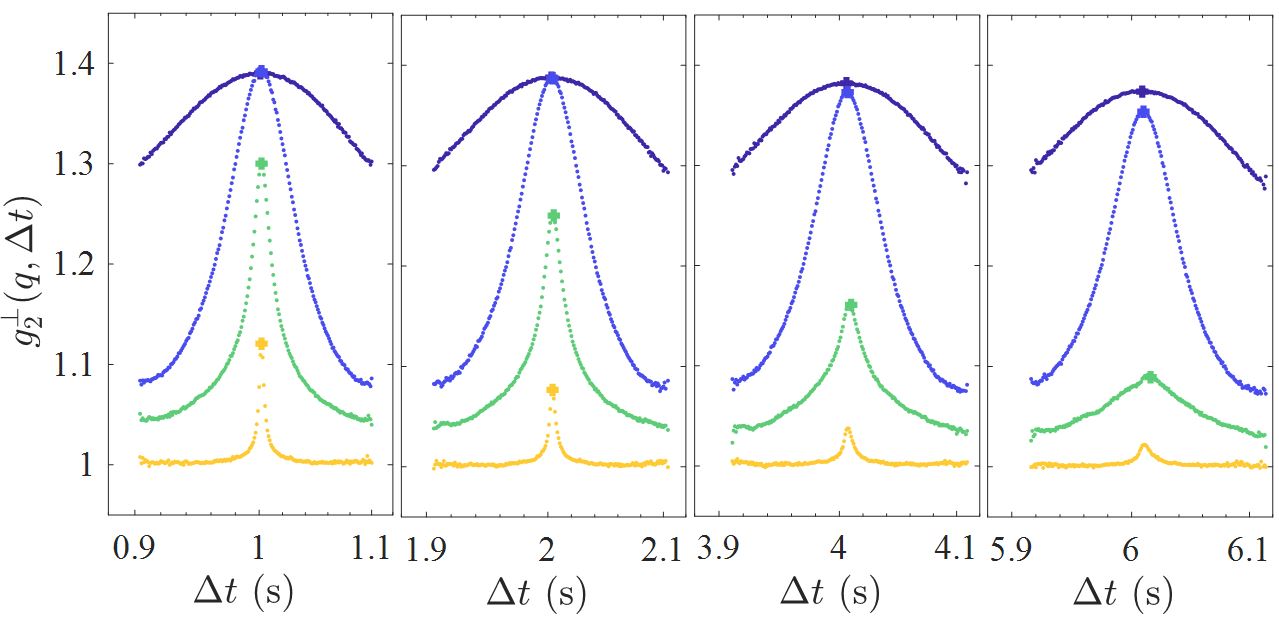}
    \caption{Echo peaks. (a) Echo peak ($g_2^{\perp}(q,\Delta t)$) at $\Delta t\simeq 1$, 2, 4, 6~s extracted from Fig.\ref{fig2:xpcs}(c). Colors from blue to yellow code for different strain amplitude, $\gamma=0.1$, 0.5, 5, 20~\% at $q=0.016$~nm$^{-1}$. 
    %(b) Half width amplitude of the echo peak $\Gamma$ as function of $q$. Evolution of the decoralation velocity $1/(\langle q\Gamma\rangle)$ as a function of $\gamma$. $\langle \rangle$ denote the average over $q$. Red lines are guides for the eyes. 
    }
    \label{figS:echo}
\end{figure}

\vspace{2mm}
\noindent\textbf{Echo correlations} -- In rheo-echo scattering experiments, the echo correlations $g_2^{\perp}(q,\Delta t)$ (Fig.~\ref{fig2:xpcs}(c)) in the vorticity direction provide two intrinsically distinct and complementary measures of microscopic dynamics: the \emph{intra-cycle shape} of individual echo peaks (Fig.~\ref{figS:echo}) and the \emph{cycle-to-cycle decay} of their amplitudes (Fig.~\ref{fig4:xpcs})(a). The temporal shape of an echo peak, characterized by its width and functional form near $\Delta t = nT$, reflects the distribution of irreversible displacements accumulated within a single oscillation cycle and should depend on $\Delta t$, $q$ and $\gamma$. A full study of the echo shape is reserved for future publication.
%Under general conditions where irreversible motion grows approximately linearly with time away from the echo center, the peak shape is governed by the statistics of local rearrangement rates, with narrow, Gaussian-like peaks indicating relatively homogeneous dynamics and broader, heavy-tailed peaks signaling increasing spatial heterogeneity and the presence of rare, large rearrangement events. 
By contrast, the amplitude of the echo peak at $\Delta t = nT$ quantifies the cumulative loss of configurational memory over many cycles. \tg{In this sense, rheo-echo-XPCS does not directly probe persistent particle trajectories but rather the progressive renewal of the microstructure under cyclic loading.} Its decay with increasing $n$ reflects the progressive accumulation of irreversible rearrangements that prevent full microstructural recovery from cycle to cycle, independent of their detailed temporal distribution within individual cycles. While the echo shape probes instantaneous heterogeneity on the timescale of a single oscillation, the echo amplitude encodes long-time correlations and collective effects governing how local rearrangements accumulate and propagate through the material over many cycles. Together, these two observables disentangle short-time, local dynamical heterogeneity from long-time, cumulative irreversibility, providing a multi-timescale framework for characterizing yielding and flow in driven soft materials.

%%%%%%%%%%%%%%%%%%%%%%%%%%%%%%%%%%%%
%%%%%%%%%%%%%%%%%%%%%%%%%%%%%%%%%%%%

%%
\vspace{2mm}
\noindent\textbf{Specificity of the non-linear acquisition scheme couple to rheology} --
The approach allows us to track up to 100 successive echoes, providing robust statistics and access to the temporal evolution of the system over many deformation cycles. At the same time, the burst acquisition significantly reduces X-ray exposure compared to a continuous (linear) acquisition scheme as well as the number of frames. For 100 deformation cycles of 1~s, XPCS acquisition is performed in burst mode with a duration of 0.2~s per cycle, resulting in a total X-ray exposure time of 20~s. By contrast, a continuous acquisition at 1~kHz would produce $200\times 10^3$ images instead of $20\times 10^3$ and correspond to a total exposure time of 100~s.

In contrast to previous rheo-echo experiments performed on gels, our measurements are $q$-resolved over nearly one decade. This extended wavevector range allows us to directly probe the $q$-dependence of the characteristic relaxation times, revealing a clear ballistic scaling. Such scaling would not be accessible without sufficient $q$ resolution and highlights the additional insight provided by our approach into the underlying microscopic dynamics and a yielding mechanism.

Thanks to the large detector distance from the sample (30~m), we access a smaller $q$ range, enabling us to probe cluster-scale structures, in contrast to more conventional setups where the detector distance is typically about 10~m. Moreover, compared to standard configurations at shorter sample–detector distances, the speckle contrast is significantly enhanced ($\beta \simeq 0.4$ rather than $\beta \simeq 0.1$), which improves the signal-to-noise ratio and allows for finer resolution of the echo peak along the $y$-direction.

Compared to \cite{rogers2014} who performed the first rheo-echo-XPCS experiment, we did not need to stop the strain at its maximum to measure the XCPS. This indicates that the rheology is optimal, allowing us to obtain an accurate measurement of the response stress, $G'$ and $G''$ during the XPCS.

In~\cite{kamani2025}, the two-time correlation function $g_2(\Delta t,t)$ is evaluated within oscillation cycles and integrated over all scattering directions during oscillatory shear experiments on colloidal gels. This intra-cycle approach connects microscopic structural memory to macroscopic rheological memory and quantifies the progressive accumulation of irreversibility during cyclic deformation. Because all sources of decorrelation contribute simultaneously, including affine advection, non-affine rearrangements, and heterogeneous yielding, the measured loss of correlation reflects a global measure of structural irreversibility within a given oscillation period.
In contrast, our rheo-echo-XPCS protocol probes correlations between successive oscillation cycles rather than within a single cycle. By measuring echoes at fixed macroscopic strain in the vorticity direction, reversible affine contributions are strongly suppressed and the measured decorrelation directly reflects the cumulative configurational renewal occurring from one cycle to the next. The two approaches therefore probe complementary temporal aspects of yielding under oscillatory shear: intra-cycle structural reversibility in~\cite{kamani2025}, and cycle-to-cycle accumulation of microscopic irreversibility in the present work.

\begin{figure}
    \centering
    \includegraphics[width=0.45\textwidth]{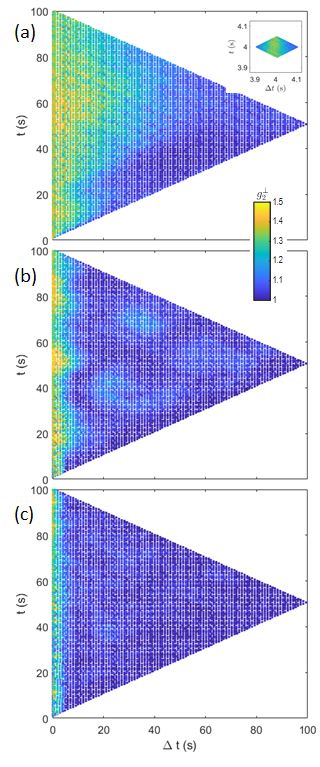}
    \caption{Two-time auto-correlation function (TTCF), $g_2(\Delta t,t)$ at $q = 0.014~\text{nm}^{-1}$ for $\gamma$=0.5 (a), 5 (b) and 10~\% (c). Note that the original full TTCF image was binned 8x8 to allow for plotting within my computer's memory limits.
    The TTCF consists of discrete lozenge-shaped measurement windows, as shown in the inset of panel (a).
    Inset in (a) is a zoom of the echo measured at $\Delta t=4$~s  and $t=$4~s. The intensity of $g_2(\Delta t,t)$ is color coded from blue (1) to yellow (1.5). No binning in the inset.}
    \label{figS:ttcf}
\end{figure}

%%%%%%%%%%%%%%%%%%
\vspace{2mm}
\noindent \textbf{Two-time auto-correlation function, $g_2(\Delta t,t)$} -- 
Fig.~\ref{figS:ttcf} displays the two-time autocorrelation function $g_2(\Delta t, t)$ measured at a given $q$ and strain amplitude $\gamma = 0.5$, 5 and 10\%. Each point-like zone corresponds to an echo peak. 
The inset in Fig.~\ref{figS:ttcf}(a) provides a zoomed view of an echo peak.

\begin{figure}[h]
    \centering
    \includegraphics[width=0.45\textwidth]{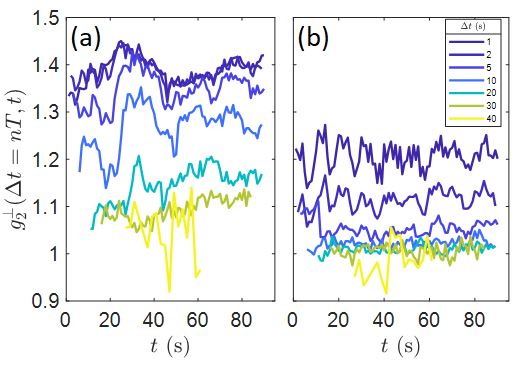}
    \caption{Echo peak amplitude as function of time $t$ for different $\Delta t=nT$ at $q = 0.014~\text{nm}^{-1}$ for  $\gamma = 0.5\%$ (a) and  $\gamma =10\%$ (b). The different $\Delta t$ are color coded from dark blue (1~s) to yellow (40~s).}
    \label{figS:sta}
\end{figure}

In Fig.~\ref{figS:sta} we show the time evolution of the amplidude of the echo peak $g_2(\Delta t,t)$  at a given $q$ for different lag time $\Delta t=nT=\{1,2,5,10,20,30,40\}\,\mathrm{s}$. 
%The data is averaged along $t$ on each individual echo to facilitate the visualation of the time evolution.
%The stationarity of the two-time correlation function was tested by evaluating the TTCF at a series of fixed lag times$ \Delta t_{\mathrm{target}}=\{1,2,5,10,20,30,40\}\,\mathrm{s}$. 
Typically, for stationary dynamics, the correlation amplitude remains independent of age, while systematic variations indicate
non-stationary or aging behavior. 
Assessing stationarity in rheo-echo-XPCS is more challenging than in conventional XPCS because the dynamics are periodically driven by the applied shear, such that the two-time correlation function is intrinsically structured by echo peaks. As a result, stationarity must be evaluated at fixed echo times and as a function of age, which significantly reduces the number of statistically independent points compared to standard stationary XPCS measurements.
%We averaged along the center of each echo peak (as displayed in inset  Fig.~\ref{figS:ttcf}(a)) with the criteria$|t-\Delta t_{\mathrm{target}}|<10^{-3}\,\mathrm{s}$. 
As shown in Fig.~\ref{figS:sta}(a-b), the resulting age-dependent correlation functions, $g_2(\Delta t_{\mathrm{target}}, t)$, do not seem to exhibit any systematic age dependence (i.e., no tilt along $t$) for either strain amplitude, $\gamma = 0.5\%$ or $\gamma = 10\%$. This observation is consistent with the rheological measurements, which indicate a constant stress during the XPCS acquisition (Fig.~\ref{figS1:sigma}(b)). \tg{These observations support the assumption of stationarity over the duration of the experiment and therefore justify the time averaging used to compute $g_2(q,\Delta t)$.}

%%%%%%%%%%%%%%%%%%%%%%%%%%%%%%%%%%%%
%%%%%%%%%%%%%%%%%%%%%%%%%%%%%%%%%%%%

%%%%%%%%%%%%%%%%%%%%%%%%%%%%%%%%%%%%
%%%%%%%%%%%%%%%%%%%%%%%%%%%%%%%%%%%%

%%

%We propose here a temptative interpretation. At low strain amplitudes ($\gamma\ll\gamma_y$), irreversible events are rare and bond‐rupture–driven elastic stress waves dominate.  These coherent, propagating rearrangements could sharpen the correlation decay leadting to $\alpha>1$. As $\gamma\to\gamma_y$, plastic rearrangements become significant and more heterogeneous.  The distribution of relaxation times broadens, driving $\alpha$ toward unity, characteristic of simple exponential decay. Finaly, beyond yielding ($\gamma>\gamma_y$), the network is extensively disrupted.  Dynamics become highly heterogeneous, with a broad spectrum of local rearrangements, leading to $\alpha<1\,$.
%In summary, the strain‐dependence of $\alpha$ in Fig.\ref{figS3:alpha} reflects the evolving balance between ballistic, elastic‐wave–mediated relaxation (compressed exponentials) and spatially heterogeneous plastic rearrangements (stretched or simple exponentials) as the gel transitions from solid‐like to yielded flow.  

%%%%%%%%%%%%%%%%%%
\vspace{2mm}
\noindent \textbf{Model for the decorrelation process} -- 
Fig.~\ref{fig4:xpcs}(e) shows, in the vorticity direction, the evolution of the stretching exponent $\alpha$ obtained by fitting the maximum of each echo‐peak positioned at $\Delta t=nT$ to $g_1(q,\Delta t)=\exp\bigl[-(\Delta t/\tau)^\alpha\bigr]$. As the strain amplitude increases, we observe that $\alpha$ decreases, going from $\sim 1.5$ to $\sim 0.5$ while $\tau$ remains ballistic. We propose here a tentative interpretation to account for this observation. 

%%%%%%%%%%%%%%%%%%%%%%%%%%%%%%%%%%%%
 \begin{figure}[h]
    \centering
    \includegraphics[width=0.45\textwidth]{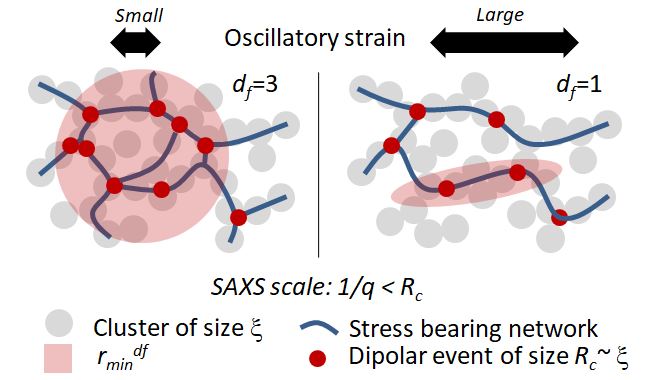}
    \caption{Sketch of the model}
    \label{figS:sketch}
\end{figure}
%%%%%%%%%%%%%%%%%%%%%%%%%%%%%%%%%%%%

We generalize the approach of Cipelletti \textit{et al.}~\cite{cipelletti2000} by introducing an effective decorrelation dimension $d_f$ (see Fig~\ref{figS:sketch}). In this framework, the field correlation function $g_1(q,\Delta t)$ is governed by rare, localized rearrangements that act as elastic dipoles embedded in the network. A dipolar event of characteristic size $R_c$ generates a far-field displacement 
\begin{equation}
u(r,\Delta t)\sim \frac{e(\Delta t)\,R_c^3}{r^2},
\end{equation}
where $e(\Delta t)$ is the accumulated local strain. We take $R_c$ to be of the order of the cluster size, since the largest inhomogeneities dominate the long-range response, and focus on the asymptotic regime $r\gg R_c$.  

Scatterers located within a distance $r<r_{\min}(\Delta t)$ from the dipole, where the condition $q\,u(r,\Delta t)\gtrsim1$ is met, undergo displacements large enough to decorrelate the scattering signal. This criterion defines the minimal distance
\begin{equation}
r_{\min}(\Delta t)\sim \left(q\,e(\Delta t)R_c^3\right)^{1/2}.
\end{equation}

If the number of scatterers affected by an event scales as $r_{\min}^{d_f}$, the survival probability is proportional to $g_1$ and reads
\begin{equation}
g_1(q,\Delta t)\simeq e^{-C\,r_{\min}(t)^{\,d_f}}
=e^{-C\,(q\,e(\Delta t)\,R_c^3)^{d_f/2}},
\end{equation}
with $C$ an effective event density and $d_f$ the dimension of stress bearing structure. In this generalized formulation, the decay of the dipolar displacement field is assumed to remain that of a three-dimensional elastic medium, i.e. $u(r)\sim r^{-2}$. What changes with decreasing $\alpha$ or $d_f$ is not the form of the displacement field itself, but the geometry of the stress-bearing backbone that transmits it. In the intact gel we recover the model developed in~\cite{cipelletti2000}, rearrangements couple to scatterers throughout a three-dimensional volume, corresponding to $d_f=3$. As $\alpha$ decreases, the stress bearing network changes dimensionality from $d_f=3$ to a lower value but remain mechanically connected. As a result, the number of scatterers coherently affected by an event scales with distance to the power $d_f$ rather than cubically, i.e, the effective event density scale as $C\sim L^{-d_f}$.

Assuming ballistic growth $e(\Delta t)=\dot\varepsilon_{\rm eff} \Delta t$, where $\dot\varepsilon_{\rm eff}$ is an effective local strain rate that increases with imposed strain amplitude/frequency, we obtain
\begin{equation}
\begin{aligned}
g_1(q,\Delta t) &= 
\exp\!\Big[-C \left( q\, \dot{\varepsilon}_{\mathrm{eff}}\, \Delta t\, R_c^3 \right)^{d_f/2} \Big] \\
&= \exp\!\Big[-\left( \frac{\Delta t}{\tau(q)} \right)^{\alpha} \Big],
\end{aligned}
\end{equation}
where
\begin{equation}
\tau(q) = \frac{1}{C^{2/d_f} \, R_c^3 \, \dot{\varepsilon}_{\mathrm{eff}} \, q}, 
\quad \alpha = \frac{d_f}{2}, \quad v_{\tau}=\frac{1}{\langle \tau q \rangle}.
\end{equation}

\noindent $\langle .\rangle$ denote the average over $q$ and incorporate the average relaxation time defined as
$\langle \tau \rangle = \frac{\tau}{\alpha} \, \Gamma\!\left(\frac{1}{\alpha}\right)$.

\vspace{2mm}
\noindent\textbf{Discussion of the model -- }
Following Eq.~7 and 8, it emerges that yielding is governed by a decorrelation velocity $v_{\tau}$ and the exponent $\alpha=d_f/2$. Let us now confront this model to our experiments. 

{First}, the model successfully captures the ballistic scaling $\tau \propto 1/q$ observed across all measured strain amplitudes as shwon in Fig.~\ref{fig4:xpcs}(b).

{Second}, we may test the dependence of $v_{\tau}$ with $\gamma$, as shown in Fig.~\ref{fig4:xpcs}(d). This is nontrivial because it relies on relating $\dot{\varepsilon}_{\mathrm{eff}}$, the local irreversible strain accumulation rate, to the macroscopic strain $\gamma$. Beyond the linear regime, we may assume $\dot{\varepsilon}_{\mathrm{eff}} \propto \gamma f$, which leads to $v_{\tau}\propto \gamma$. In Fig.~\ref{fig4:xpcs}(d), we observe $v_{\tau}\propto \gamma^{0.8}$. The slight deviation from linear scaling ($\gamma^{0.8}$ vs $\gamma^{1.0}$) could reflect additional nonlinear effects in the strain accumulation.

{Third}, we focus on the relation $\alpha = d_f / 2$. For an intact gel, scatterers are coupled through a three-dimensional elastic network, corresponding to $d_f = 3$ and $\alpha = 1.5$. One then recovers a compressed exponential with $\alpha = 1.5$, in agreement with the original model~\cite{cipelletti2000}. This scenario is observed at low strains, in the linear regime, where strain does not induce irreversible changes. It remains valid up to $\gamma \simeq 0.5\%$, as shown in Fig.~\ref{fig4:xpcs}(d,e), where $\tan \delta$, $v_{\tau}$, and $\alpha$ are quasi-independent of $\gamma$.
At higher strains, $\alpha$ decreases from 1.5 to roughly 1 with increasing strain \tg{amplitude} (Fig.~\ref{fig4:xpcs}(e)). Within the model framework, this behavior indicates a reduction of the effective dimension $d_f$ from 3 to 1. Under strain, stress transmission becomes confined to quasi-one-dimensional filaments rather than the full 3D network, as sketched in Fig.~\ref{figS:sketch}. While local displacements remain ballistic ($e(t) \propto t$), decorrelation shifts from volume-based to filamentary processes. Self-consistency further requires $q^{-1} \ll L$, with $L$ the filament length, so that $r_{\min}(\tau) \ll L$.
Although this stress-bearing backbone is not evident in the static scattering $I(q)$, simulations of colloidal gels under increasing strain amplitude show stress localization into filamentous structures~\cite{colombo2014}, supporting our hypothesis that $d_f$ may decrease to 1 to account for the observed evolution of $\alpha$. We note, however, that we do not observe strain hardening prior to yielding, as reported in~\cite{colombo2014}.

Fourth, we focus on the interpretation of the decorrelation velocity $v_\tau = 1/\langle q\tau\rangle$. It emerges naturally from the ballistic scaling $\tau \propto q^{-1}$ and serves as a characteristic velocity scale governing structural memory loss in the gel. It is important to clarify what $v_\tau$ represents physically: it is \emph{not} the literal speed at which individual particles move during rearrangements, but rather the rate at which structural decorrelation propagates through the material following irreversible events.
To understand this distinction, consider the mechanism underlying ballistic relaxation in colloidal gels. When an inter-cluster bond ruptures, it releases localized stress that propagates through the elastic network as stress redistribution or elastic waves~\cite{cipelletti2000}. This propagation causes particles throughout a correlated volume to undergo small displacements. The scattering correlation decays when these accumulated displacements satisfy $q \cdot u(r,t) \gtrsim 1$, where $u(r,t)$ is the displacement field. For a dipolar rearrangement with characteristic size $R_c$ creating a far-field displacement $u(r,t) \sim \varepsilon(t) R_c^3/r^2$, where $\varepsilon(t)$ is the accumulated local strain, the decorrelation occurs when particles at distance $r_{\text{min}} \sim (q\varepsilon R_c^3)^{1/2}$ are affected. The number of decorrelated scatterers scales as $r_{\text{min}}^{d_f}$, leading to correlation decay on a timescale $\tau$ set by the rate of strain accumulation.
With ballistic growth of irreversible strain, $\varepsilon(t) \sim \dot{\varepsilon}_{\text{eff}} t$, dimensional analysis yields $\tau \sim 1/(q v_\tau)$, where $v_\tau \sim \dot{\varepsilon}_{\text{eff}} R_c$. Thus, $v_\tau$ quantifies the velocity scale at which the decorrelation front---the boundary beyond which scatterers have lost correlation with their initial positions---expands through the material. In this interpretation, $v_\tau$ is fundamentally a measure of how rapidly the gel "forgets" its initial structure due to accumulated plastic rearrangements. \tg{The observed $q^{-1}$ dependence should not be interpreted as evidence for persistent directed particle trajectories across oscillation cycles. Rather, it reflects the strain-like character of the irreversible decorrelation process measured by rheo-echo-XPCS, arising from elastically correlated rearrangements and progressive configurational renewal.} The linear relationship $v_\tau \propto \tan\delta$ (Fig.~\ref{fig4:xpcs}d) then reveals that the macroscopic loss tangent, directly controls the rate of this microscopic memory loss across all deformation regimes.
The intermediate scattering function is measured exclusively for scattering vectors oriented along the vorticity direction; accordingly, the extracted relaxation time $\tau(q)$, stretching exponent $\alpha$, and effective decorrelation dimension $d_f$ should be understood as directionally projected quantities. The model does not assume isotropic relaxation dynamics but instead describes how irreversible rearrangements couple to density fluctuations along a direction free of affine shear motion. Although stress transmission and particle displacements are expected to be anisotropic under shear, the vorticity direction provides a privileged probe of non-affine, irreversible dynamics. Within this framework, the observed decrease of $\alpha$ reflects a reduction in the effective dimensionality of stress-bearing pathways contributing to vorticity-projected decorrelation, without implying isotropy of the full displacement field. Accordingly, fitting $g_{2}^{\perp}(q,\Delta t)$ with the present model is mathematically consistent, keeping in mind that the extracted parameters are interpreted as effective, direction-dependent quantities.
%\end{comment}
\end{document}